# Lattice small polarons and magnetic interactions drive preferential nanocrystal growth in silicon doped hematite

Mattia Allieta*[1], Marcello Marelli[1], Mauro Coduri[2,3], Mariana Stefan[4], Daniela Ghica[4], Giorgio Morello[5], Francesco Malara[1], Alberto Naldoni[6]*

[1]CNR, Istituto di Scienze e Tecnologie Molecolari (ISTM), via C. Golgi 19, 20133 Milano, Italy;

[2]ESRF- The European Synchrotron. 81, Avenue des Martyrs, 38000, Grenoble, France;

[3]University of Pavia, Department of Chemistry, via Taramelli 16, 27100, Pavia, Italy;

[4]National Institute of Materials Physics, Atomistilor 405A street, 077125 Magurele, Romania;

[5]Inorganic Chemistry Laboratory, Department of Chemistry, University of Oxford, South Parks Road, Oxford, OX13QR (UK)

[6]Regional Centre of Advanced Technologies and Materials, Faculty of Science, PalackýUniversity Olomouc, 17. listopadu 1192/12, 771 46 Olomouc, Czech Republic

## Abstract

Understanding the interplay between the structural, chemical and physical properties of nanomaterials is crucial for designing new devices with enhanced performance. In this regards, doping of metal oxides is a general strategy to tune size, morphology, charge, lattice, orbital and spin degrees of freedoms and has been shown to affect nanomaterials properties for photoelectrochemical water splitting, batteries, catalysis, magnetic applications and optics. Here we report the role of lattice small polaron in driving the morphological transition from nearly isotropic to nanowire crystals in Si doped hematite ($\alpha$-$Fe_2O_3$). Lattice small polaron formation is well evidenced by the increase of hexagonal strain and degree of distortion of $FeO_6$ showing a hyperbolic trend with increasing Si content. Local analysis via pair distribution function highlights an unreported crossover from small to large polarons, which affects the correlation length of the polaronic distortion from short to average scales. Ferromagnetic double exchange interactions between $Fe^{2+}/Fe^{3+}$ species is found to be the driving force of the crossover, constraining the chaining of chemical bonds along the [110] crystallographic direction. This promotes the increase in the reticular density of Fe atoms along the hematite basal plane only, which boosts the anisotropic growth of nanocrystals with more extended [110] facets. Our results show that magnetic and electronic interactions drive preferential crystallographic growth in doped metal oxides, thus providing a new route to design their functional properties.

Corresponding author: mattia.allieta@gmail.com, alberto.naldoni@upol.cz

## I. INTRODUCTION

Morphology is one of the most important key parameters to control the efficiency of performing nanomaterials. [1]-[6] Photocatalytic [5], electronic and magnetic devices [2],[7] are only few examples of applications, which are strictly dependent on the nanoparticles (NP) arrangement since their relative shape defines the orientation of the active nanocrystal facets and the consequent electronic and magnetic transport properties.

Recently, the role of morphology has been well evidenced in the performance of doped hematite ($\alpha$-$Fe_2O_3$),which has triggered a great interest as visible light harvesting catalysts in photoelectrochemical (PEC) splitting of water [3],[5],[6],[8]-[11]. In particular, the application of PEC water splitting on a large scale requires cheap, abundant, stable and nontoxic materials. $\alpha$-$Fe_2O_3$ fulfils all these requirements but the low electrical conductivity, the short hole diffusion length and the fast electron-hole recombination limit its PEC performance. N-type doping with aliovalent elements such as Si, enhances by orders of magnitude the conductivity of doped $\alpha$-$Fe_2O_3$, while nanostructuring improves light-trapping by modulating the nanoparticles (NPs) to a size comparable to that of hole diffusion [5],[8]-[11]. The accepted conduction mechanism in doped $\alpha$-$Fe_2O_3$ involves the formation of small polarons induced by $Fe^{2+}$species which localize the carriers to specific crystallographic sites until the thermal energy can activate the electron hopping through $Fe^{2+}$/$Fe^{3+}$.[12]-[14] Because of the spin coupling pattern imposed by antiferromagnetism (AFM) parallel to the crystal *c* axis, the additional electron can only be transferred to Fe sites within the same basal layer giving rise to an anisotropic conductivity, which has been reported to be four times greater along the [001] direction [8],[12]. Hence, the conductivity of doped $\alpha$-$Fe_2O_3$ is strongly influenced by the morphology and NPs preferentially grown along the [110] crystallographic direction have shown indeed an improved PEC activity.[8]-[11]

The mechanism of doping inducing morphological transitions has been recently discussed for Cu and Al doped $\alpha$-$Fe_2O_3$ [15],[16]. Jahn-Teller effect and anisotropic change of reticular density were invoked to explain the evolution of particle shape for $Cu^{2+}$ [15] and $Al^{3+}$ [16], respectively, where the valence of $Fe^{3+}$ resulted to be inlaterated in both cases. However, to the best of our knowledge, no study has been reported on the relationship between NPs morphology and the $Fe^{2+}$ formation providing the direct observation of the structural extention of lattice polaron from local to long range scales. In this regards, Fig.1(a) shows that the undoped hematite structure is based on the hexagonal close packed (hcp) arrangement of $O^{2-}$ anions, s.g. R-3c, $Fe^{3+}$ cations occupy the two-thirds of the octahedral sites forming $FeO_6$ units in the (001) planes. Further, $FeO_6$ octahedra share one face and three edges with their neighbors generating three short face sharing Fe-O and one Fe-Fe distances, namely $(Fe\text{-}O)_F$ and $(Fe\text{-}Fe)_F$, respectively. The long edge sharing distances are comprised of the remaining three $(Fe\text{-}O)_E$ and $(Fe\text{-}Fe)_E$ distances. First principle calculations suggested that the substitutional doping is favored with respect to interstitial doping predicting the relationship between the local expansion of the $(Fe\text{-}O)_E$ immediately connected to dopant species [see Fig.1(*b*)] and the small polaron formation [12],[13] . The presence of lattice polarons can be then inferred directly from the local evolution of Fe-O distances as a function of Si content.

In this work, we present a systematic study on Si-doped $\alpha$-$Fe_2O_3$ nanostructures investigating a wide range of Si doping. Using synchrotron powder diffraction and X-ray total scattering method based on pair distribution function (PDF), we provide a detailed description of the effect of Si doping into the long-range and local order structure of $\alpha$-$Fe_2O_3$. Our analysis is supported by a combination of trasmission electron microscope (TEM) and electron spin resonance (ESR) measurements . We found no evidence of symmetry breaking of $\alpha$-$Fe_2O_3$ space group despite the increase of Si content produces an average distortion of the lattice parameters and Fe-Fe, Fe-O bond distances consistent with the $Fe^{3+}$ valence reduction as predicted by theoretical calculations [12],[13]. PDF analysis provides evidence of local structure inhomognenities in all the investigated samples. In particular, the observed average distortion is fully riconducible to a local polaronic deformation of $FeO_6$ octahedra, which increases its magnitude toward the long-range scale with increasing doping level. We observe that Si induces the transition from NPs with ellipsoidal shape to nanowires growing preferentially along the [110] direction that self-assemble in layered superstructures. We demonstrate that the morphological transition is driven by the magnetic interactions induced by the lattice polarons describing the fundamental mechanism that controls the crystal morphology of metal oxides affected by polaronic distortion.

## II. EXPERIMENTAL TECHNIQUES AND METHODS

$\alpha$-$Fe_2O_3$ nanomaterial were prepared through a hydrothermal route following a modified procedure reported by Wang et *al.* [17]. Equal amounts of water and ethanol were mixed with $Fe(CH_3COO)_3$ and heated at $T$ = 150°C for at least 15 h in autoclave and subsequently sintered at 550 °C for 1 h. The doped samples were prepared by the same synthesis route and by adding tetramethoxysilane (TMOS) to obtain different Si content corresponding to atomic Si/(Fe+Si) = 0.22%, 0.44%, 0.60%, 0.80%, 1%, 2%, 3%, 4%, 5%, 6%, 7%, 10%, 15%, 17%, 20%, 50%.

We performed high resolution synchrotron powder diffraction experiments at the ID22 beamline of the European Synchrotron Radiation Facility (ESRF). Patterns for 0%, 1%, 2%, 3%, 4%, 5%, 6%, 7%, 10%, 15%, 17%, 20%, 50% Si-doped samples were collected at room temperature using incident photon beam with λ=0.30988 Å with the crystal analyzers' setup. Data for PDF analysis were collected using a 2D detector (Perkin Elmer XRD 1611CP3) at ID22 and λ = 0.16102 Å ($Q_{max}$=28 Å$^{-1}$) for selected samples 0%, 1%, 2%, 3%, 4%, 5%, 10%, 15%, 20%, 50% composition. In addition, samples with 0.22%, 0.44%, 0.60%, 0.80% content were acquired at room temperature on a Huber 4-axes diffractometer at the powder diffraction beamline (MCX) of the synchrotron Elettra using incident photon beam with λ = 1.03300 Å.

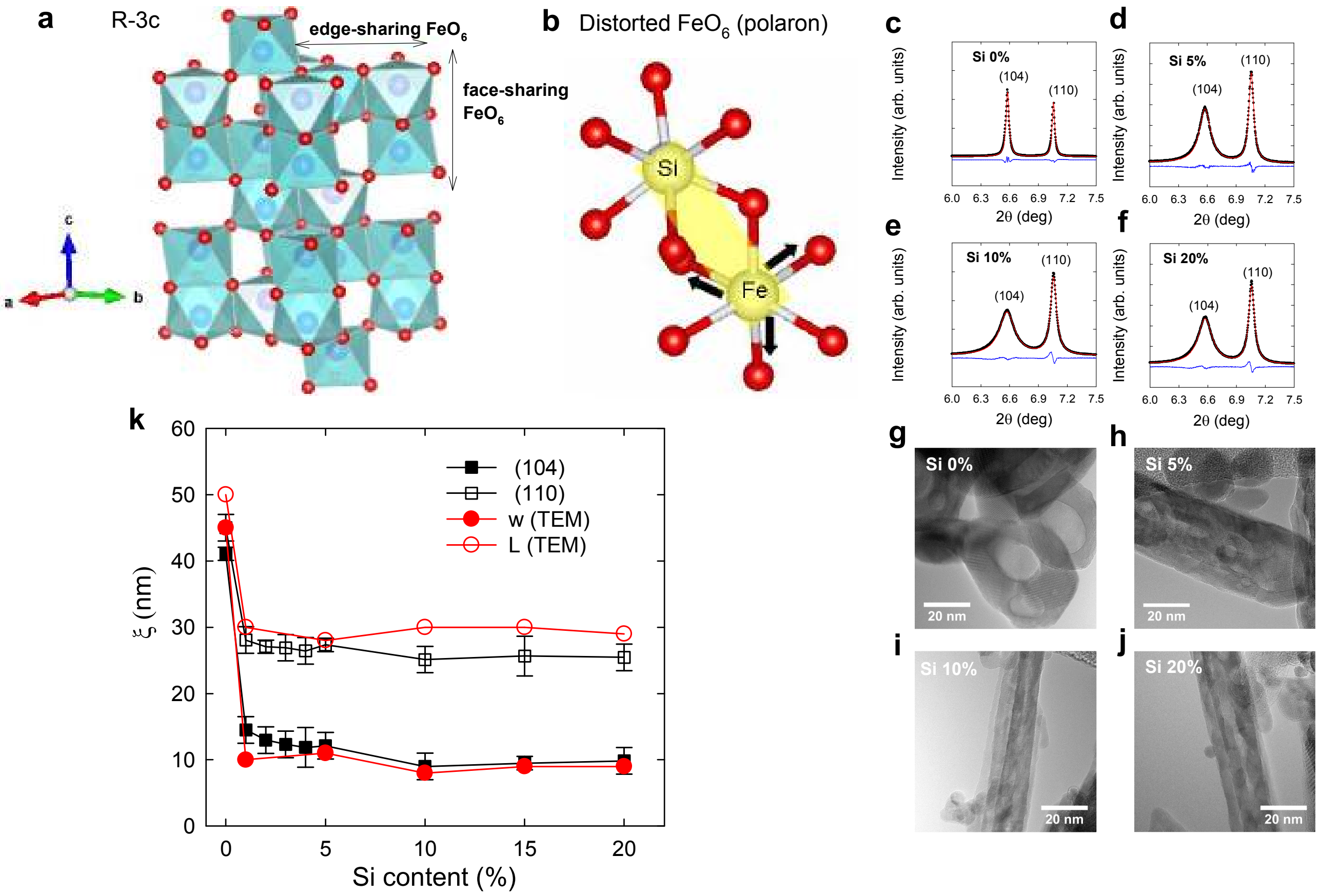

FIG.1. (Color online) (a) Representation of α-$Fe_2O_3$ unit cell (R-3c) in hexagonal setting. Arrangement of irons forming pairs of octahedra sharing edges with three neighbouring octahedral within the same (001) plane and one face with an octahedron in an adjacent plane along the *c*-axis. (b) Sketch of local $FeO_6$ octahedral environment in the presence of substitional doping. Polaron formation results in distortions associated to expansion of long $(Fe\text{-}O)_E$ distances (full arrows), respectively. (c)– (f) Region of the X-ray diffraction patterns showing the strongest (110), (104) reflections reported for selected Si content values. Dots are experimental data; continuous lines are the profiles calculated through Rietveld refinements; peaks are labelled with the respective Miller indexes. Rietveld agreement factors [$R(F2)$] between observed and calculated patterns ranged from 0.06 to 0.08. (g)-(j) In the right side of each panels, couple of TEM and HRTEM micrographs highlighting the particles shape related to the selected Si content are also shown. (k) Comparison between the coherence length ξ related to (110), (104) diffraction peaks (empty and full squares) and the nanoparticle widths (w), length (L) observed by TEM (empty and full circles). Because of strong aggregation of nanoparticles, at Si ≥ 5 % the TEM parameters are related to the single observed crystallite within the resolution of our determination.

Rietveld analysis was performed through the GSAS program [18]. We use the hematite rombohedral model of R-3c space group in hexagonal setting with Fe in 12*c* position (0,0,*z*) $z \sim 0.35$ and O in 18*e* position (x, 0, 0.25) $x \sim 0.31$ even to analyze the data of doped sample. Within this model, any contribution of Si to hematite structural evolution upon doping is accommodated in the refined parameter of the undoped structural model. During the refinement the background was subtracted using shifted Chebyshev polynomials and the diffraction peak profiles were fitted with a modified pseudo-Voigt function [18]. In the last calculation cycles all the parameters were refined: cell parameters, atomic positional degrees of freedom, isotropic thermal parameters, background, diffractometer zero point, and line profile parameters.

Selected diffraction data were reduced to the PDF using the formalism of the $G(r)$ function as implemented in the PDFGETX2 program [19]. $G(r)$ is obtained from the total structure factor S(Q) via the sine Fourier transform (FT):

$$G(r) = \frac{2}{\pi} \int_{Q=0}^{Q_{\max}} Q[S(Q)-1]\sin(Qr)dQ\,, \qquad (1)$$

where $Q_{\max} = 4\pi\sin\theta / \lambda$ and $r$ is the interatomic distance. $S(Q)$ is the experimental coherent X-ray scattering intensity after correcting the raw data for sample self-absorption, for multiple scattering, and for Compton scattering.

Structure refinements against the $G(r)$ curves were carried out using the PDFGUI program [20]. The program assesses the accuracy of the refinement by the agreement factor ($R_w$) defined as follows [20]:

$$Rw = \left[ \frac{\sum w_i (G_i^{\exp} - G_i^{calc})^2}{\sum w_i (G_i^{\exp})^2} \right]^{1/2} \qquad (2)$$

High Resolution Transmission Electron Microscopy (HRTEM) analysis were performed on selected samples by a JEOL JEM ARM 200F. The specimens were manually smashed in an agate mortar, suspended in isopropanol and sonicated 10 minutes. Each suspension were dropped onto a holey carbon coated copper TEM grid and dried overnight before the analysis.

Magnetic characterization was made by performing room temperature ESR measurements in the X-band (9.8 GHz) using a Bruker ELEXSYS-E580 spectrometer equipped with a calibrated X-band Super High QE (SHQE) cylindrical cavity resonator (ER4123SHQE). The derivative $dP/dH$ of the microwave power $P$ absorbed by the sample was recorded as a function of the static magnetic field $H$. The deconvolution and lineshape simulation of the spectra were performed with the EasySpin v. 5.1.3 program [21].

## II. RESULTS AND DISCUSSION

According to high resolution XRPD, all the samples are well described by a single phase of hematite $\alpha$-$Fe_2O_3$ rhombohedral model. Within the resolution of our measurements, we did not detect any spurious crystalline phase at each Si concentration level [10], [22].

Figure 1 (*c*)-(*f*) shows the doping evolution of (110) and (400) hematite peaks for selected Si content values. We observed a clear signature of strong anisotropic diffraction peak broadening as evidenced by the integrated intensities ratio $I_{(110)}/ I_{(400)}$ that evolves from 0% to 20 %, respectively [10]. In addition, the peak broadening is accompaniend by the increase of a background modulation at low diffraction angles with increasing the Si content [Fig.S1]. This sort of bump distributed in a wide range of $2\theta$ is generally ascribed to the presence of amorphous phase due to dopant segregation at high doping level. Here we decouple this two phenomena by firstly investigating the nature of the doping driven peak broadening and then by analysing the amorphous phase evolution via local range PDF analysis.

A comparison between XRD patterns and TEM micrographs [Fig. 1 (*g*)-(*l*), Fig.S2] well evidence the relation among Si content, XRD broadening and shape anisotropy. On average, the morphology of the undoped sample is well described by a nearly pseudo-spherical particle with a width (*w*) of ~ 45 nm and length (*L*) of ~ 50 nm. The crystallite pseudo-isotropic shape is reflected by the diffraction line profile for which we observe nearly absent anisotropic broadening. After the introduction of Si 1 %, the nanocrystals changed into hollow ellipsoids with *w* ~ 10 nm and *L* ~ 30 nm and nanocrystal wall ~ 3 nm thick. With increasing Si content, a marked shape evolution is even more evident [10]. At Si 5 %, $\alpha$-$Fe_2O_3$ nanostructures mainly consist in nanowires and microfiber-like building blocks [Fig.1(*h*)] that self-assemble through electrostatic interactions into bundles with morphological habit similar to common fibrous silicates [23]. All these samples showed a full solid structure, demonstrating that at dopant concentration higher than 1 %, the incorporation of silicon into the $\alpha$-$Fe_2O_3$ structure drives directly the growth of nanowires with eventually higher hierarchical organization. In particular, the morphological transition is gradual as Si 10% is composed of a mixture of nanowires having *w* ~ 8 nm and some rounded nanoparticles. On the other hand, Si 20% sample contains large aggregates of ~ 10 nm thick nanowires assembled along their longer lateral dimension and preserve the same average morphology observed at lower doping level [ Fig.1(*h*),(*i*)]. At high Si content, the silicon segregation appears clear and detectable along with the fibrous morphology [Fig. S2, Fig.S3].

In Fig. 1 (*k*), we compare *w, L* values extracted from TEM micrographs with the average coherence length ($\xi$/nm) calculated from XRPD $2\theta$ and FWHM of (104), (110) peaks. We note that for the undoped hematite $\xi(110) \sim \xi$ (104), in agreement with the nearly isotropic particle shape observed by TEM. At Si 1 % we determined average $\xi(110) \sim 30$ nm and $\xi(104) \sim 10$ nm which are in good agreement with *w, L* TEM values. This correspondence which is almost Si content independent [Fig. 1 (*k*)], defines a relationship between the observed average ellipsoidal shape and the anisotropic peak broadening in Si doped $\alpha$-$Fe_2O_3$. In particular, it is well known that the scattering related to (110), (104) is highly sensitive to preferential growth of hematite crystal planes [8],[10],[11]. The narrow profile of (110) consistent with the more extended length of ellipsoidal nanoparticles provides indication of a preferential crystal growth along the [110] crystallographic direction [10],[11]. The morphological transition from pseudo-spherical to nanowires particles is driven by the effect of Si elogating on average the crystal shape along the [110] direction. Since in hematite the ferromagnetic interactions favor dramatically the electron conduction along the [110] direction [8],[10],[11], our hypothesis is that the introduction of Si constraints the crystal growth by strengheting the charge/magnetic interaction channels between Fe species along this crystallographic direction.

In order to assess this hitherto unreported interplay between charge, spin, lattice degrees of freedom and particle morphology, we start investigating the structural contribution of Si on $\alpha$-$Fe_2O_3$ by Rietveld refinements. In Fig.2(a)-(c) we report the doping evolution of all the refined structural parameters. Hexagonal strain is defined as $\eta = c/a$ where $a$ and $c$ are the hexagonal unit cell axes. At 0% Si $\eta$ = 2.731, in agreement with the reported value for bulk hematite [24]. With increasing Si content we observed a hyperbolic growth of $\eta$ saturating above Si ~ 4%. This evolution can be even better appreciated considering doping dependent degree of distortion of $FeO_6$ octahedra ($D$) for edge $(Fe\text{-}O)_E$ and face sharing $(Fe\text{-}O)_F$ distances [Fig.2(b)], which is defined as follows [25]:

$$D = \frac{(Fe - O)_E}{(Fe - O)_F} - 1 \qquad (3)$$

Figure 2(d) shows the order parameter $D$ as a function of Si content. The parameter follows the same trend as $\eta$ and it is driven by the expansion of $(Fe\text{-}O)_E$ and the contraction of $(Fe\text{-}O)_F$ distances [Fig.2(b)]. Similarly, $(Fe\text{-}Fe)_F$ and $(Fe\text{-}Fe)_E$ exhibit a similar contraction with increasing Si saturating at the same solubility limit detected for other parameters [Fig.2(c)]. The observed doping dependence of selected order parameters highlights the effect that the introduction of Si produced on hematite structure.

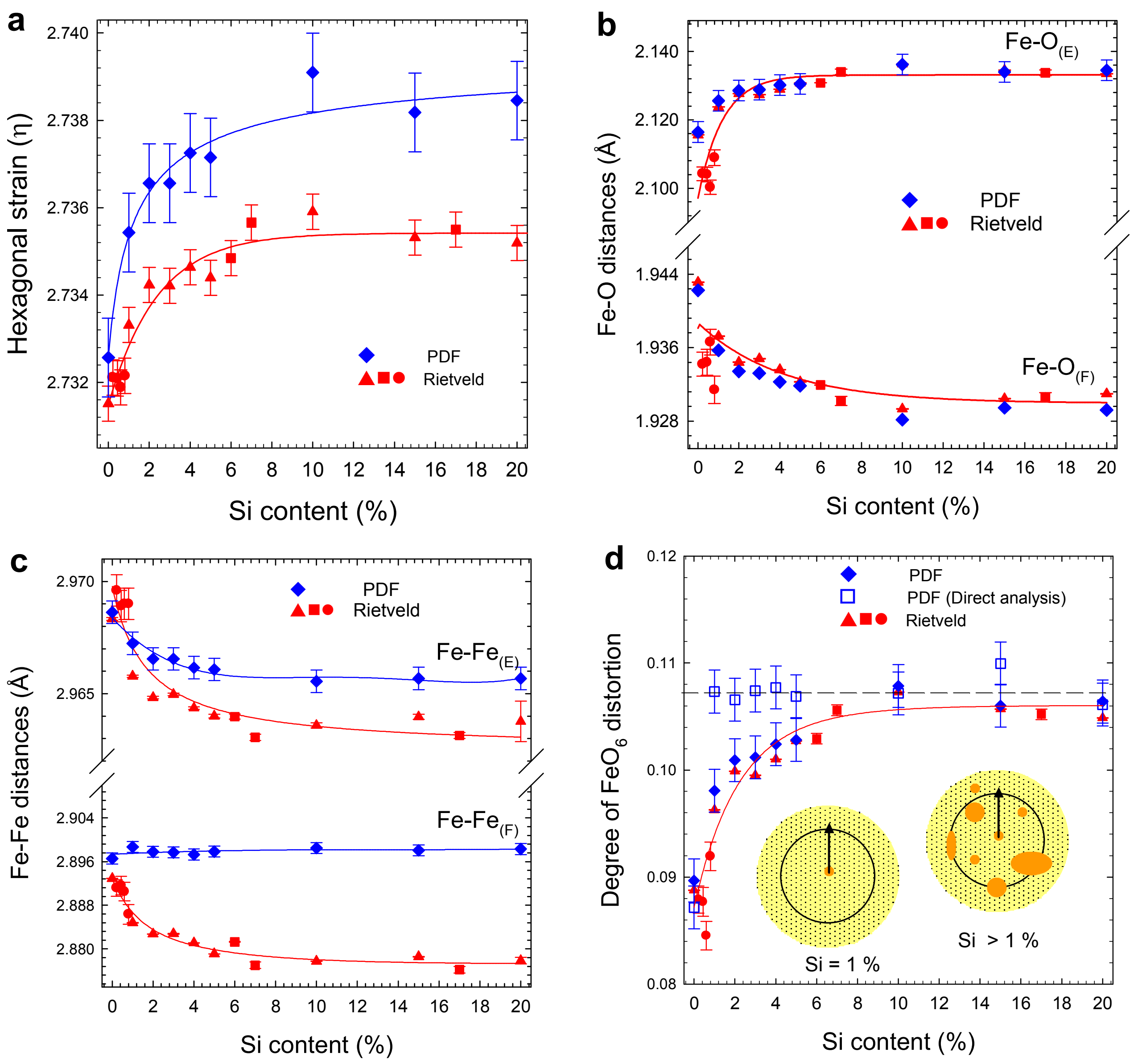

FIG.2 (Color online) (*a*) Hexagonal strain expressed as $\eta = c/a$ as a function of Si content (%). (b),(c) Doping evolution of edge and face shared Fe-O, Fe-Fe distances, respectively. (d) Degree of $FeO_6$ distortion ($D$) as a function of Si content (%). (d) Degree of $FeO_6$ distortion calculated by using the Fe-O distances extracted by PDF direct analysis ($D_{DA}$). Inset: 2D illustrations of the corrisponding atom-atom pairs probed by the PDF containing the sample region distorted by polarons (full spots) embedded in the undistorted hematite structure (dotted area) are shown. The origin of the arrow in each illustrations concides with the Fe-O first shell whereas the circle encloses the interatomic distances ($r$) up to ~ 15 Å. In all the panels full triangles, squares and circles are data obtained by Rietveld refinements, while full diamonds are the results obtained by full structure profile PDF refinement in the selected local $r$ range.

Substitutional doping was widely reported both theoretically [12],[13] and experimentally [10] to be the main mechanism for this solid solution. However, the details about the interplay and the interactions of the possibile atomic defects generated upon doping are still lacking to date. Here, to figure out the role played by Si in the structural modification of hematite, we refer to the substitutional model expressed by the following equation:

$$2Fe_{Fe}^{x} + 3O_{o}^{x} + ySi \rightarrow (2-2y)Fe_{Fe}^{x} + ySi_{Fe}^{\bullet} + yFe_{Fe}^{'} + 3O_{O}^{x} \quad (4)$$

where, in Kruger-Vink notation [26], $Fe_{Fe}^{x}$ and $O_{o}^{x}$ are species located at the sites of undoped hematite, while $Si_{Fe}^{\bullet}$ and $Fe_{Fe}^{'}$ are $Si^{4+}$ and $Fe^{2+}$ species located at the sites of doped hematite.

The substitution of Fe by Si generates a positive charged defect $Si_{Fe}^{\bullet}$ that is compensated by the reduction of iron valence via $Fe_{Fe}^{'}$ formation. Because of the smaller ionic radius (*ir*) of $Si^{4+}$ (*ir* = 0.40 Å) than $Fe^{3+}$ (*ir* = 0.65 Å), one could expect a contraction of $FeO_6$ unit. However the concomitant formation of bigger $Fe^{2+}$ (*ir* = 0.78 Å) gives rise to an expansion of $FeO_6$ due to deformation of Fe-O immediately conneted to dopant species. This is directly supported by the evolution of D parameter providing a direct evidence of the $Fe^{2+}$ distortion associated to small polaron formation prevailing over the structural effect due to Si *ir* only. Moreover, the lengthening of $Si_{Fe}^{\bullet}$-O-$Fe_{Fe}^{'}$ implies the shrinkage of the distances of adjacent $Fe_{Fe}^{'}$-O-$Fe_{Fe}^{x}$-O-$Fe_{Fe}^{x}$ chains which well agrees with the contraction of both $(Fe\text{-}Fe)_F$ and $(Fe\text{-}Fe)_E$ [Fig.2(c)]. Doping dependence of hexagonal strain is finally fully consistent with the deformation of $FeO_6$ octahedra providing evidence that $\eta$, $D$ are reliable order parameters to quantify the magnitude of structural distortion due to small polaron formation. Analogous expansion of $FeO_6$ octahedra was considered a clear experimental evidence of the polaronic distortion even in Fe doped $LiNbO_3$ crystals [27].

The substitutional doping model is consistent with our Rietveld results and the observed structural modifications are likely to be caused by the competing distortion modes as determined by the interaction of atomic defects [eq.(4)]. Within the small polaron picture, the introduced electrons are localized on Fe sites polarizing neighboring atoms and deforming the surrounding lattice. Hence, the polaron radius is small and the deformation is confined at the local scale being of the order of about one unit cell. Our order parameter $D$ is obtained by the analysis of diffraction data related to the reciprocal space and hence can be associated to the long range/average structure. As a result, the polaronic deformation of $FeO_6$ must imply some kind of interplay between the short and long range structures, requiring a more complex picture to reconcile the observed average distortions with the emergence of local polarons.

We now turn our attention to the short range structure of Si doped $\alpha$-$Fe_2O_3$. Figure S4 shows the experimental PDF obtained in the $1 \leq r \leq 10$ Å range for undoped and Si 1% samples. Each positive peak in $G(r)$ function is proportional to the probability of finding two atoms separated by a distance $r$ averaged over all pairs of atoms in the sample. Remarkable differences between undoped and doped sample are evident by comparing the PDF peaks. M-shape features in the difference curve $\Delta G(r)$ (continuous line at the bottom of Fig.S4) clearly reflect peak broadening together with additional positional mismatch typical of appreciable changes in the underlying bond-length distribution. Both effects are consistent with an enhancement of the static disorder which is fully compatible with the observed $FeO_6$ deformation.

To analyze the local structure evolution more quantitatively, the experimental PDF were fitted in the $1.7 \leq r \leq 12$ Å range using the structural models employed for interpreting diffraction patterns. To keep the analysis simple, again we used the undoped hematite structural model by monitoring the evolution of the agreement factor, $R_W$ [eq.(2)] together with the refined structural parameters.

A careful inspection of all the PDF against $\alpha$-$Fe_2O_3$ hexagonal model did not reveal additional peaks above ~2 Å indicating in principle the inability of determining Si-based phases even using PDF. Nevertheless, both the emergence of low diffraction angle modulation at high Si content and the saturation of all structural order parameters suggest the segregation of doped Si somehow. Given the contrast between the X-ray scattering factors of the element pairs involved, i.e., Fe-Fe, Fe-O, Fe-Si, Si-Si, Si-O, the partial $g_{Si\text{-}O}(r)$ and $g_{Si\text{-}Si}(r)$ have the lowest weighting. Then, the PDF peaks related to the Si-O, Si-Si pairs might correspond to very weak features underneath the main peaks even at high Si content. In order to increase the weight of Si partial g($r$), we analyzed the PDF collected on Si 50% sample that indeed displays an additional peak at around 1.61 Å [inset to Fig.S1]. This feature is consistent with Si-O first shell distances supporting that Si segregates into an amorphous $SiO_2$ phase above ~ 4 % . Having identified the nature of the amorphous phase, in Fig.S5 we report the short range PDF refinements performed for undoped and Si 1% samples. As expected, the undoped sample shows full agreement with the R-3c model with local refined degree of distortion of $FeO_6$ [$D_{PDF}$ =0.085(3)] which is in good agreement with the value determined from the reciprocal space analysis [$D$=0.083(1)]. Conversely, for 1 % sample the refined PDF systematically underestimates the intensity of the experimental peaks in the first ~ 4 Å [difference curve in Fig.S5(b)], providing evidence that the structural parameters obtained in the $1.7 \leq r \leq 12$ Å are inadequate to describe the very short range. Since in the small polaron scenario the lattice deformation extends over the length of one unit cell, we decided to extract long and short Fe-O distances with two different approaches: (1) by fitting directly Gaussian functions to the FeO first shell, i.e. direct analysis (DA); (2) by calculating the distances values from the full profile least squares refinement in the $2.7 \leq r \leq 12$ Å range. Figure 3 (a)-(h) shows the calculated profiles determined from both direct analysis (DA) [Fig.3(a)-(d)] and PDF refinements [Fig.3(e)-(h)] for selected compositions.

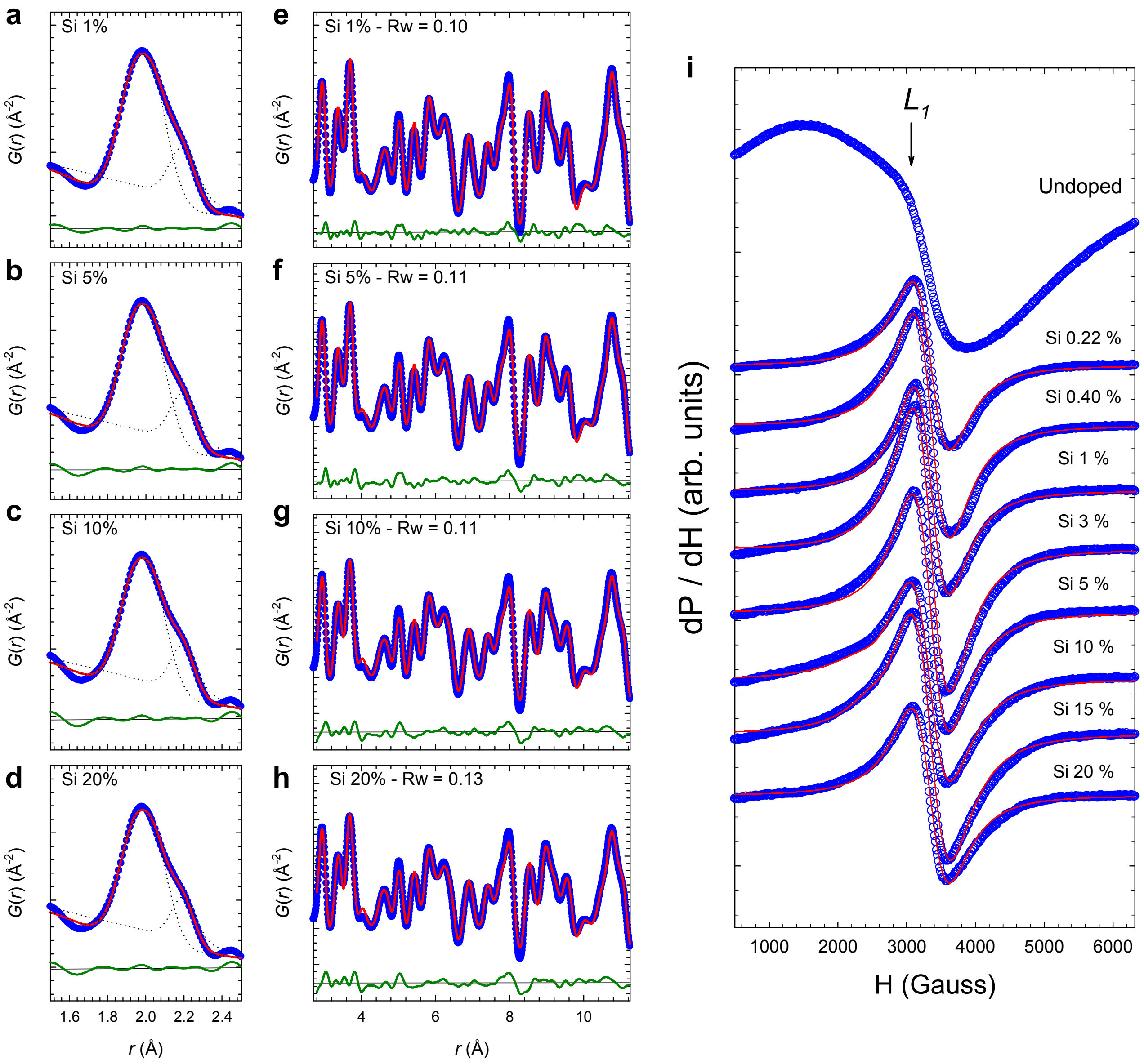

FIG.3 (Color online) (*a*) – (*d*) Direct PDF analysis related to nearest-neighbor Fe-O as a function of selected Si content (%).(*e*) – (*h*) Full structure profile PDF refinement for the same sample composition obtained by using the rhombohedral average model in a selected *r* range excluding the Fe-O first shells. Dots and continuous lines are observed and calculated PDF data, respectively. The residuals are also shown at the bottom of each figures. (*i*) Doping evolution of ESR spectra for selected samples. Dots are observed data whereas continuous are calculated Lorentzian spectra.

We note that by excluding the first Fe-O shell from the PDF refinement ($r > 2.7$ Å), for all samples the hexagonal R-3c model gives a reasonable fit over the selected *r* range. In this context the local deformation of $FeO_6$ induced by Si ($r$ < 2.7 Å) is rationalized in terms of magnitude of distortion mismatch which can be quantified by the degree of $FeO_6$ distortion [eq. (4)] determined both using Fe-O values obtained from direct analysis ($D_{DA}$) and from the results of PDF fitting ($D_{PDF}$). These values are compared in Fig.2(d) with the respective average values ($D$) obtained by Rietveld analysis. On the short-medium range scale (2.7 ≤ $r$ ≤ 12 Å) the evolution of $D_{PDF}$ follows the $D$ behavior matching the solubility limit observed on average. On the other hand, the $D_{DA}$ values are independent of composition with the magnitude of the $FeO_6$ distortion consistent with saturation values of $D$ and $D_{PDF}$ observed for Si > 4 %. This behavior is comparable to that reported for other mixed valence systems, e.g. $La_{1-x}Sr_xMnO_3$, where the magnitude of the local Mn-O distances (~2.1 Å) related to Jahn-Teller small polarons was found to be unchanged by Sr doping [28]. Given the very small radius of polaron distortion, the $D_{DA}$ can be considered a direct measurement of the structural effect induced by the electron localization. According to the observed doping evolution of $D$ parameters, it is most likely that the orientation and the magnitude of the local small polaron varies from site to site since the $D_{PDF}$ values match the average distortion in a wide $r$ range below the solubility limit. This site orientational disorder is even corroborated by the dependence of $\eta$ at the local scale [Fig.2(a)]. In particular, the behavior is consistent with the fluctuation of the primary order parameter among the differently strained sample regions resulting in a higher local value of $\eta$ than long scale, as observed for all the doped sample investigated. In addition, we noted that locally the Fe-Fe distances display a different evolution with doping in comparison to average scale [Fig.2(c)]. Specifically, edge sharing $(Fe\text{-}Fe)_E$ distances decrease in a similar manner as observed at longer scale but the face sharing $(Fe\text{-}Fe)_F$ are almost independent from the composition. Since face and edge sharing $FeO_6$ are parallel to $c$-axis and $ab$ plane, respectively, the local evolution of the $(Fe\text{-}Fe)_F$ distances reveals that the contribution of small polaron via $Fe^{2+}$ formation affect the structure more deeply on $ab$ plane. The value of this observation will be reveal below as its relationship with the preferred crystal growth direction is central to explain the morphological transition in this system.

To illustrate how the Si doping induces the polaronic distortion from local to long range structures, we sketch a 2D illustration of two selected ranges of composition probed by XRD and PDF techniques namely (i) Si = 1 % and (ii) Si > 1% [Fig.2(d)]. Every sketch represents a portion of the undistorted average $\alpha$-$Fe_2O_3$ structure whereas the full spots are the regions affected by the polaronic distortion. In this view the interplay between Si doping and small polaron formation can be razionalized as follows. At Si = 1%, because of $Fe^{3+}$ valence reduction, the $FeO_6$ are distorted in their immediate surrounding. The corresponding $D_{DA}$ = 0.107(2) reflects the absolute magnitude of polaronic distortion achieved in this system whereas the great discrepancy with $D_{DPF}$ obtained at $r$ > 2.71 Å provides a direct evidence of the small character of $Fe'_{Fe}$ polarons. For Si>1 % more extended sample regions start to be distorted and the interaction between polarons is testified by the fact that locally the $D_{DA}$ remains unchanged but in the medium scale $D_{PDF}$ follows the averagebehavior. In other words, with increasing Si content the correlation length of the polaronic distortion, quantified by $D_{DA}$, becomes progressively less spatially limited promoting a transition from local to long range collective polarons interactions. At the solubility limit Si ~ 4%, $D_{DA}$ ~ $D_{PDF}$ ~ $D$ supports a complete crossover from small to large structural polarons. In this regime, the degree of distortion of $FeO_6$ is roughly the same from very local to average scales and then the structure turns to be less affected by the atomic fluctuactions recovering homogeneity. Above the solubility limit, all structural order parameters ($\eta$, $D_{DA}$, $D_{PDF}$, $D$) eventually saturate and no more doping contributes to the overall structural distortion.

Given the AFM ground state of $\alpha$-$Fe_2O_3$, a possible driving force for the observed effect could be the magnetic ordering eventually perturbed by Si. To validate this hypothesis, we studied the evolution of the ESR signal as a function of doping at room temperature. ESR allows direct access to the spin-environment interactions of Fe ions in different valence states [29], [30].

Fig.3(*i*) shows the room temperature ESR spectrum of undoped α-$Fe_2O_3$ . The signal shows a "two-line pattern" which is typical for superparamagnetic resonance observed in other Fe-based nanomaterials [31]. The spectrum is composed by a broader signal superimposed on a narrow one, with their relative intensity that depends both on the particle size and the shape distribution. The more intense and narrow line, marked as $L_1$ in Fig.3(*i*), exhibits a resonance signal at g ≈ 2.122 typical of $Fe^{3+}$ ions with $3d^5$ configuration and $^6S_{5/2}$ ground state. It is well known that the $Fe^{2+}$ ions are ESR silent at room temperature in X-band due to their very large zero-field splitting, therefore the ESR signal in doped samples originates from the $Fe^{3+}$ ions as well, as a shift or/and modulation of their resonance $L_1$.

By Si-doping, the ESR signal of α-$Fe_2O_3$ indeed dramatically changes [Fig.3(*i*)]. They are better described by a single Lorentzian line where the main parameters can be determined by fitting the spectra with the following shape function [21]:

$$\frac{dP}{dH} = \frac{d}{dH}\left[\frac{2A}{\pi} \times \left(\frac{\Delta H_{FWHM}}{4(H-H_r)^2 + \Delta H_{FWHM}^2}\right)\right] \qquad (5)$$

Where A is the area under the ESR curve, i.e. integrated signal intensity, $H_r$ is the resonance field and $\Delta H_{FWHM}$ is the full width at half maximum. The peak to peak linewidth ($\Delta H_{pp}$) is derived from $\Delta H_{FWHM}$ using the relation $\Delta H_{pp} = \Delta H_{FWHM} / \sqrt{3}$ . The fit results at selected compositions are shown by the solid lines in Fig.3(*i*).

The doping dependence of the linewidth $\Delta H_{pp}$ and the Lorentzian ESR line intensity is shown in Fig.4(a) and Fig.S6, respectively. A sudden narrowing of the $L_1$ resonance with a drop of ~ 1000 G is already observed at 0.22% Si content. On the other hand, above 0.22% the $\Delta H_{pp}$ linewidth increases and the signal broadening is accompanied by a more prominent increase of the ESR intensity (Fig.S4). Above ~ 4% both these parameters saturate up to 20% Si content.

Our undoped sample displays a deviation from the theoretical $g$ value for the free $Fe^{3+}$ ion ($g$ = 2.0023) that is generally assigned to $Fe^{3+}$ clusters that interact through superexchange (SE) coupling [31]. More precisely, as depicted in Fig.4(b), the main SE interactions in α-$Fe_2O_3$ are represented by two couplings: (i) $Fe^{3+}$-O-$Fe^{3+}$ channel sited in the same layer [marked as A in Fig.4(b)]; (ii) $Fe^{3+}$-O-$Fe^{3+}$ channel sited in adjacent layers [marked as A and B in Fig.4(b)]. By introducing $Fe^{2+}$ ions in the A layer a double exchange (DE) between mixed valence $Fe^{3+}/Fe^{2+}$ ions occurs. In DE mechanism, the electron of the $t_{2g}$ orbitals [$Fe^{2+}$ in Fig.4(b)] can hop into an empty $t_{2g}$ orbitals on the second ion ($Fe^{3+}$) and, following Hund's rule, the spin of the itinerant electron is parallel to the electron spin in the second ion leading to ferromagnetic (FM) coupling [31]. In the case of a strong isotropic exchange interaction, like SE, the Gaussian ESR line which depends on the magnetic dipole interaction ($H_d$) is narrowed into a Lorentzian line by fast dynamics resulting from the exchange coupling ($J$) according to $\Delta H_{pp} \approx H_d^2 / \omega_{ex}$ where $\omega_{ex} \approx J/\hbar$, i.e. *exchange narrowing* [29].

On the other hand, the DE hetero-spin dipolar interaction gives rise to an effective alternating FM internal magnetic field that relaxes the $Fe^{3+}$ spins [31]. Upon doping one could expect an increase of the spin relaxation rate of $Fe^{3+}$ which produces the broadening of the ESR signal, i.e. *exchange broadening* [29]. Therefore, the final linewidth $\Delta H_{PP}$ depends on the relative strengths of these two mechanisms.

The increase of the $\Delta H_{PP}$ and the Lorentzian ESR line intensity observed above 0.22% is in agreement with the DE *exchange broadening* mechanism, providing ~~a~~ further evidence of the increase of the $Fe^{2+}$ species concentration as the Si content is increased. On the other hand, the $\Delta H_{PP}$ drop of the L1 resonance in the undoped sample may be explained by the enhancement of the frequency $\omega_{ex}$ due to SE interactions. However, it should be noted that this latter mechanism is only valid for homo-spin dipolar interactions whereas the linewidth drop at 0.22% is clearly caused by the presence of $Fe^{2+}$ species.

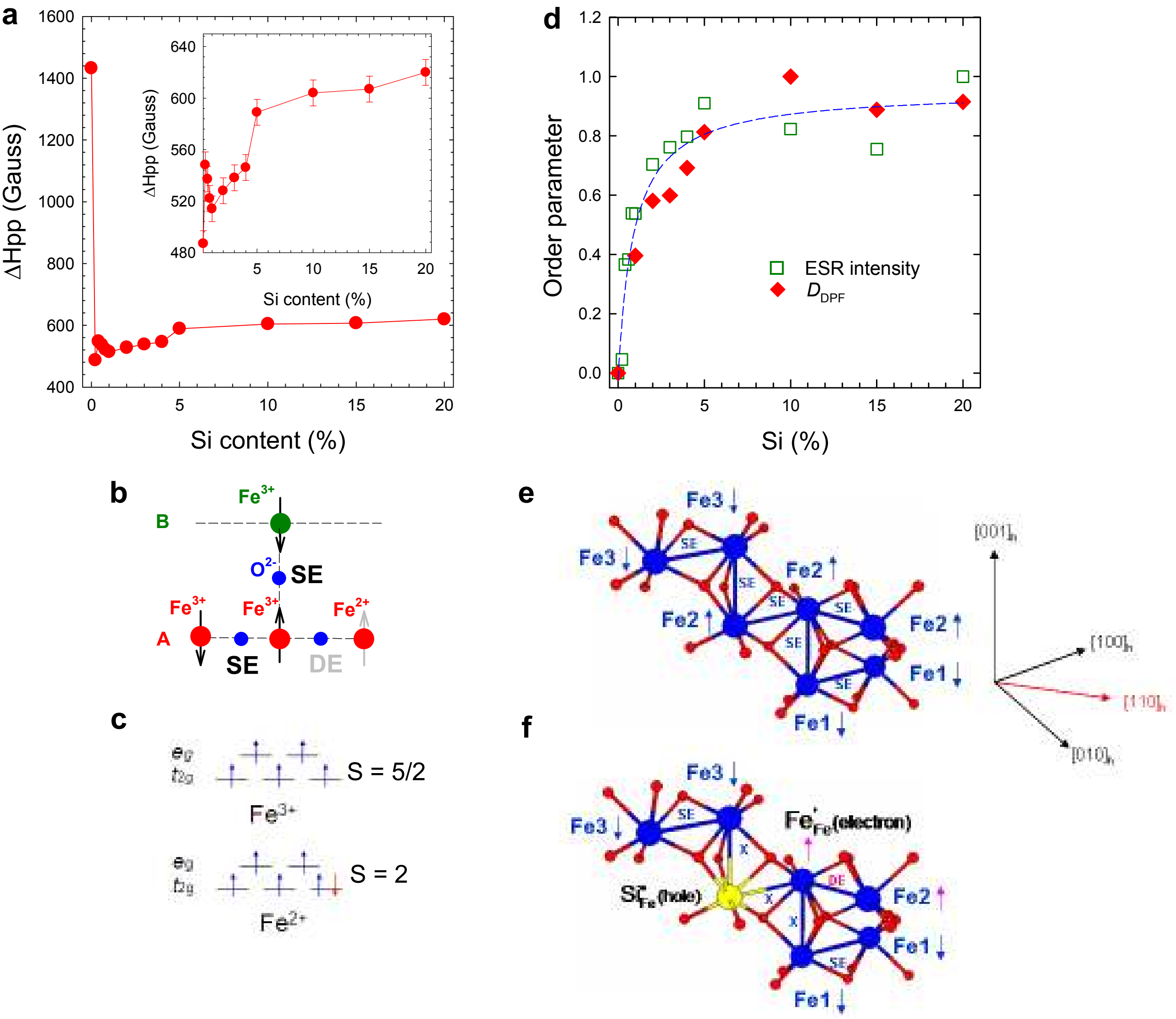

FIG.4 (Color online) (*a*) Linewidth ($\Delta H_{pp}$) as a function of Si content. Inset: detailed view of the $\Delta H_{pp}$ variation for the 0.22-20% Si content. (*b*) Superexchange (SE) between $Fe^{3+}$ ions and double exchange (DE) between $Fe^{3+}$ and $Fe^{2+}$ ions belonging to layer *A* and *B*. (*c*) Electronic structure of $Fe^{3+}$ and $Fe^{2+}$ ions in $FeO_6$ environment involving DE of $t_{2g}$ electron. (*d*) Doping evolution of normalized $D_{PDF}$ as obtained from PDF (squares) and normalized integrated ESR intensity (diamonds) (*e*), (f) Cluster models for crystallite growth mechanism in undoped and Si doped hematite, respectively. Labelling on Fe1, Fe2, Fe3 are referred to different layers stacked along the $[001]_h$ direction. Spin orientations are reported on each cations and magnetic interactions between atoms are indicated by SE = superexchange, DE = double exchange and x = null.

Looking for other possibile mechanisms, we recall that in a similar doping range, i.e. 0.1 – 1 %, n-type elements such as Ti, Si, Ge, Sn decreased the resistivity of $\alpha$-$Fe_2O_3$ by orders of magnitude, e.g. in undoped $\alpha$-$Fe_2O_3$ $\rho = 6.5 \cdot 10^5\ \Omega$ cm, while in 0.1% Ti doped $\alpha$-$Fe_2O_3$ $\rho = 5.1\ \Omega$ cm at room temperature. It is generally accepted that in $\alpha$-$Fe_2O_3$ the electron mobility increases by the motion of small polarons which involves the hopping of $t_{2g}$ electrons with its associated spin from one site ($Fe_{Fe}^{x}$) to another site ($Fe_{Fe}^{'}$). This leads to the random motion of the magnetic moments, that in turn, may lead to "motional narrowing", with the motional frequencies comparable to the strength of the exchange broadening interaction. The observed L1 linewidth drop [Fig.4(a)] can be qualitatively explained with the motional narrowing as a result of polaron mobility and it suggests a linkage with the compositional independent degree of local polaronic distortion of $FeO_6$ quantified by $D_{DA}$ [Fig.3(d)]. In this perspective, the single narrower Lorentzian resonance observed from 0.22 to 20 % Si-doping suggests that locally the maximum degree of polaronic distortion is already achieved at low Si content and it is independent from the composition. From a qualitative point of view, the motional narrowing mechanism allows us to establish a direct interplay between the polarons diffusion/interactions suggested by the $\Delta H_{pp}$ and the polaronic lattice distortion mapped by $D_{DA}$. This relationship is even better outlined by comparing the integrated ESR intensity and the order parameter extracted from PDF ($D_{PDF}$) normalized to unity for the doped samples [Fig.4(d)]. It is clear that the normalized ESR intensity and $D_{PDF}$ feature fairly equivalent doping dependences, both saturating above the solubility limit, i.e. Si ~ 4%. This provides that the polaronic distortion of $FeO_6$ is related to the $Fe^{3+}$-O-$Fe^{2+}$ spin-spin DE exchange via strong spin-lattice interaction. The ESR intensity is generally proportional to the population difference between the two states involved in the allowed transitions ($\Delta M = \pm 1$) reflecting the ESR active centers in the sample. In our case, according to exchange broadening, the parameter describes how the DE interactions strength with increasing Si. On the other hand, the proposed crossover from local to long range lattice polaron is explained by the growing of small range polaronic distorted regions which interact as more $Fe_{Fe}^{'}$ are induced by Si up to the solubility limit. We can then account for the spin-lattice coupling by considering that the doping-induced small polarons aggregate with each other because of the strong collective interactions of the DE paths driving the observed $\Delta H_{pp}$ and ESR intensity enhancement. We can then visualize the Si doped $\alpha$-$Fe_2O_3$ as an inhomogeneous pattern where, depending on the doping level, small polarons are held together by DE heterospin interactions embedded in the $\alpha$-$Fe_2O_3$ matrix dominated by SE interactions.

The structural findings reported here clarify some peculiar aspects of the morphological transition featured by Si doped $\alpha$-$Fe_2O_3$. Figure 4(e) shows a cluster model of undoped $\alpha$-$Fe_2O_3$ representing the coupling of $Fe_{Fe}^{x}$ within three different layers, i.e Fe1, Fe2, Fe3. At room temperature, Fe ions are AFM coupled by SE interactions across the shared octahedral faces along the [001] direction. In contrast, in the *ab* plane there are two interpenetrating AFM sublattices assembled by spins with a canting angle of $< 0.1°$ resulting in a weak FM interaction. As observed in our ESR measurements, adding one $Si_{Fe}^{\bullet}$ changes dramatically the magnetic interactions in the above cluster [Fig. 4(*f*)]. The induced $Fe_{Fe}^{'}$ polaron interacts ferromagnetically with first neighbors $Fe_{Fe}^{x}$ via DE exchange, whereas the introduction of Si reduces the density of SE coupling because of the increase of non-magnetic Si-O-Fe interactions. Lower dimensionality of magnetic interactions together with itinerant electrons which can be only transferred within the same layer (DE exchange along [001] would require forbidden spin flip) constrain polarons to interact only along [110] [Fig.4(*f*)]. This provides a clear evidence to define a relationship between the small polaron formation and morphological transition. The observation of a compositionally independent small polaron distortion at the local scale is common for other strongly electron systems such as doped manganites [28], [32]. Nanostructuring acts as a further degree of freedom introducing a spatial confinement that strongly affects the correlation length of the polaronic distortion. The role of small polaron, during the formation of gel-like network mediated by tetramethoxysilane (TMOS), is hence reflected by the more extended length of nanowires along the [110] since this is the crystallographic direction along which polaron are constrained to interact according to the model proposed in Fig.4(f). In terms of Bravais law, during the nanocrystal growth, the formation of small polaron constrained along the [110] direction affects the difference between reticular densities of atoms which in turns causes the difference in the growth rate of prominent crystal faces [16]. Local evolution of Fe-Fe distances is clearly sensitive to this balance. Distance contraction is only observed indeed for edge sharing distances indicating that the Fe reticular density increases barely along the *ab* plane. This leads to a slow growth of the basal faces resulting in nanocrystals with more extended [110] direction reflected, hence, by the nanowire shape.

## III. CONCLUSIONS

In summary, we demonstrated that the morphological transition from nanoparticles to nanowires in silicon doped hematite is driven by the magnetic interactions induced by the lattice polarons. We have shown that Si substitutes Fe in hematite up to a solubility limit of about 4% above which the dopant segregates as a $SiO_2$-like amorphous phase. We observed polaronic distortion due to $Fe^{3+}$ valence reduction in agreement with the distortion mode predicted by theoretical calculations. Our order parameter $D$ quantifies this effect providing a clear evidence of the average structure evolution affected by small polaron formation. However, the structural distortion at the average scale cannot be accounted by the short range character of small polaron entities. PDF analysis showed that the local structure of all the investigated doped samples displays an enhanced magnitude of $FeO_6$ distortion which is independent from the composition. The mismatch between the magnitude of $FeO_6$ distortion at short and at long ranges was then reconciled by considering a crossover from small to large polaron regimes driven by the progressive aggregation between polarons upon doping. Spin-spin interactions studied by ESR corroborated this scenario as the doping evolution of the linewidth $\Delta H_{pp}$ is well explained by considering the combination of polarons motional narrowing and exchange broadening due to ferromagnetic DE strengthening with increasing Si amount.

The findings reported here provide atomistic insights into the interplay between magnetic, electronic and preferential nanocrystals growth in doped metal oxides opening new routes to design their functional properties.

**IV. ACKNOWLEDGEMENT**

Authors would gratefully acknowledge the CERIC consortium for support through TA access to the Laboratory of Atomic Structures and Defects in Advanced Materials (LASDAM –National Institute of Material Physics, Bucharest – Romania), proposal 20162015, for the HRTEM analysis and Dr. Maraloiu for the precious support Financial support from the Romanian Ministry for Research and Innovation through the NIMP Core Program 21N/2019 is gratefully acknowledged. A.N. acknowledge the support by the Operational Programme Research, Development and Education - European Regional Development Fund, project no. CZ.02.1.01/0.0/0.0/15_003/0000416 of the Ministry of Education, Youth and Sports of the Czech Republic.

## V. SUPPORTING INFORMATION

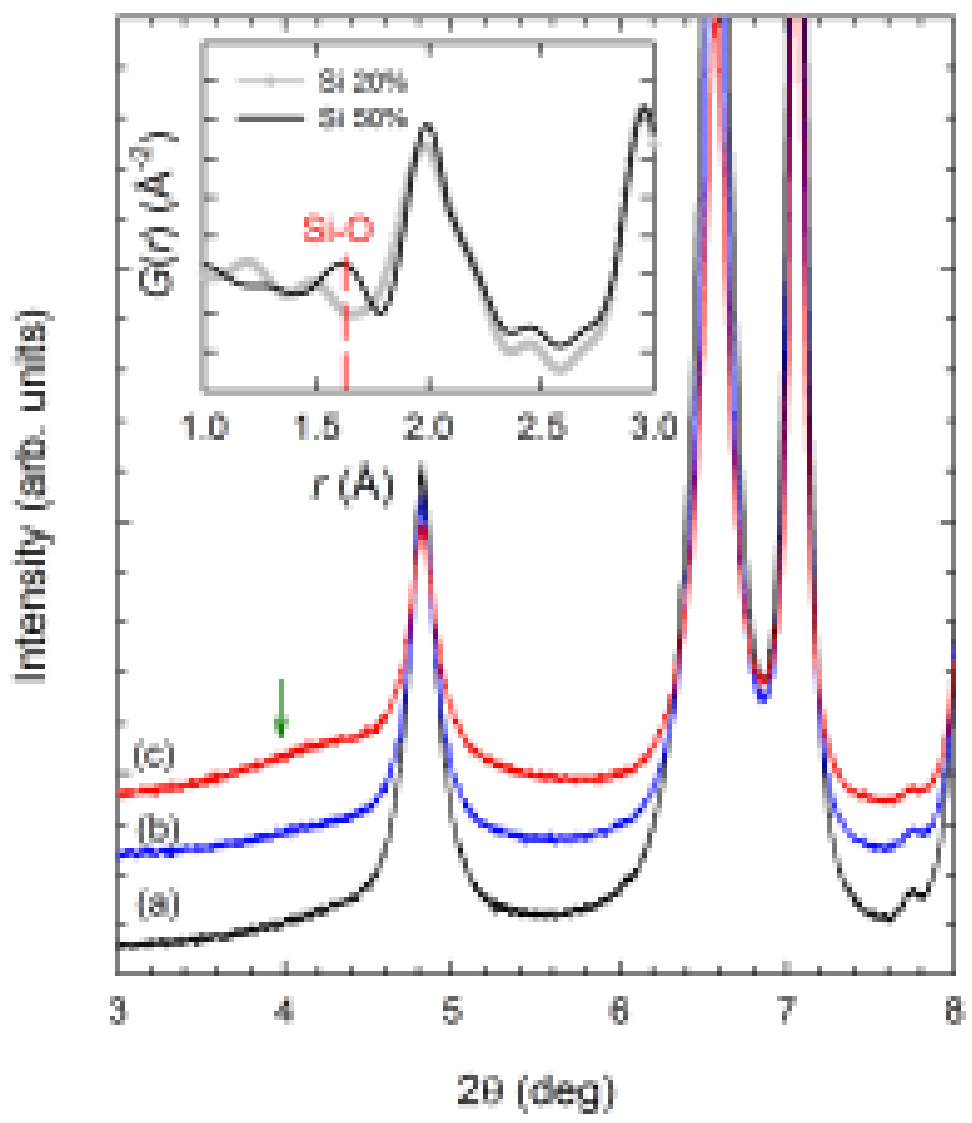

FIG.S1 (Color online) Low angle region of the X ray diffraction patterns related to Si 10% (a), Si 20 % (b) and Si 50% (c). In the insert short range PDF highlights the peak compatible to Si-O distances found in Si 50% sample only.

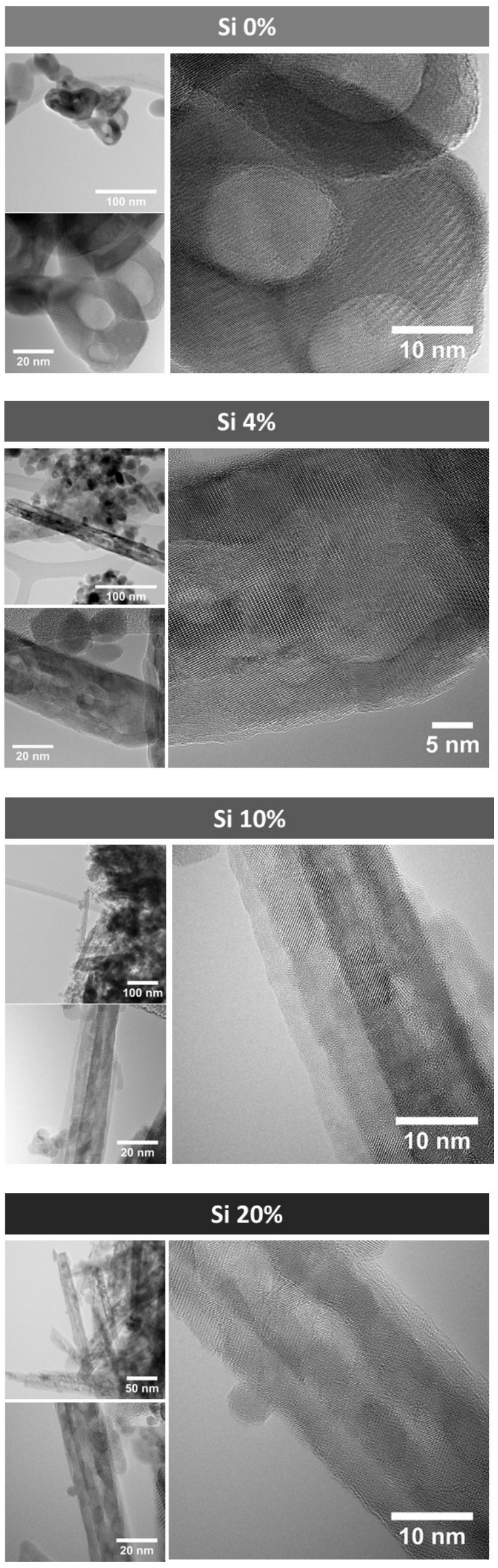

FIG.S2 TEM and  HRTEM micrographs for selected samples with a Si doping loading of 0-4-10-20

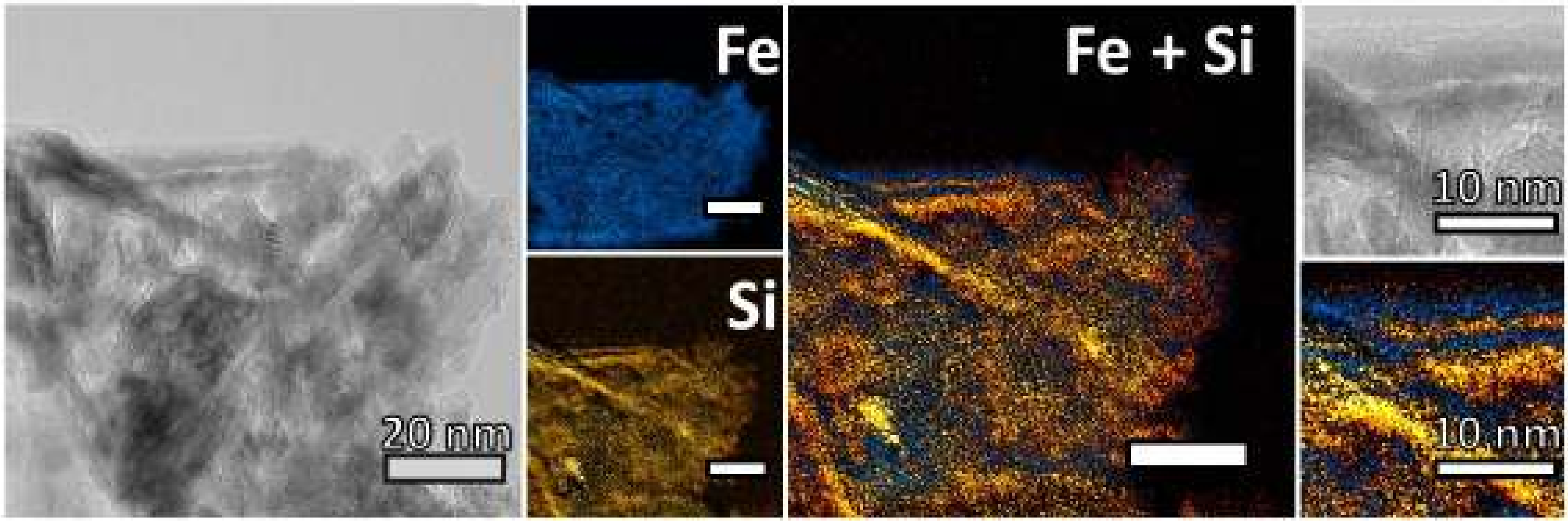

FIG S3: Elemental mapping by electron spectroscopic imaging (ESI) collecting signals at L edges for both Fe and Si for the 50% Si doped hematite

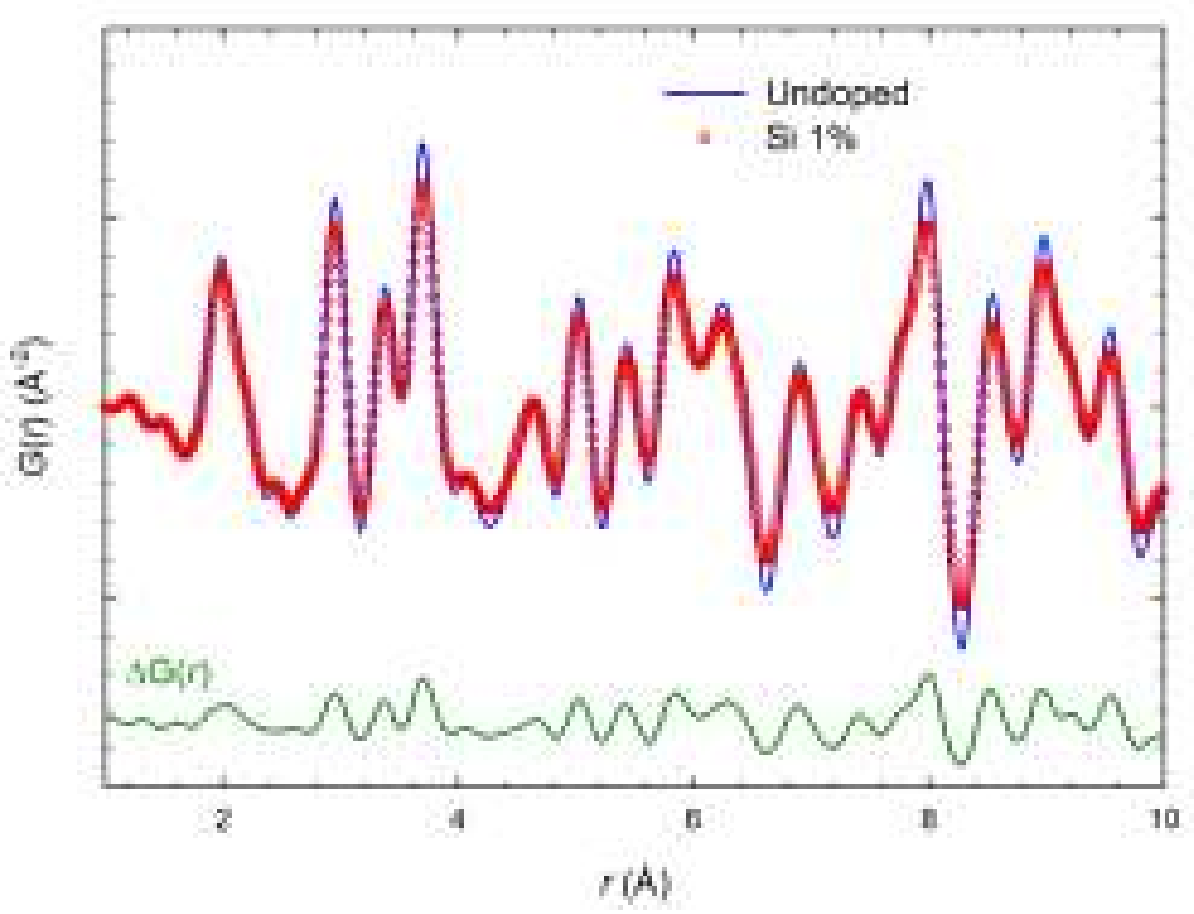

FIG. S4 (Color online) PDFs obtained from undoped (continuons lines) andm Si 1% (open circles) doped $Fe_2O_3$ within $r \sim 10$ Å . $\Delta G(r)$ is the difference between PDFs shown at the bottom of the panel.

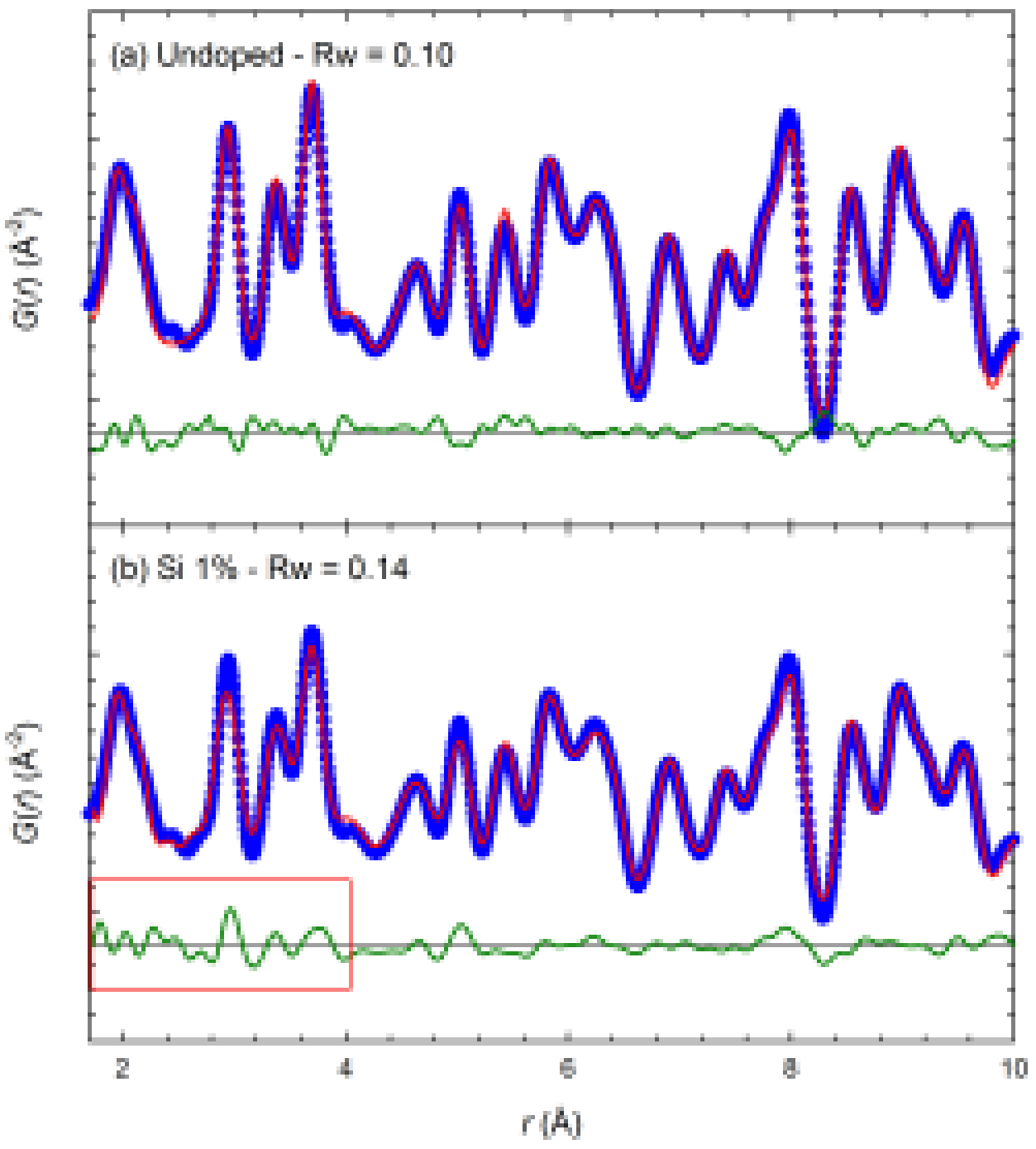

FIG.S5 (Color online) (*a*), (b) Full structure profile PDF refinement for undoped and 1% Si samples obtained by using the rombohedral average model in a selected $r$ range from $1.7 \leq r \leq 12$ Å . Dots and continous lines are observed and calculated PDF data, respectively. The residuals are also shown at the bottom of each figures

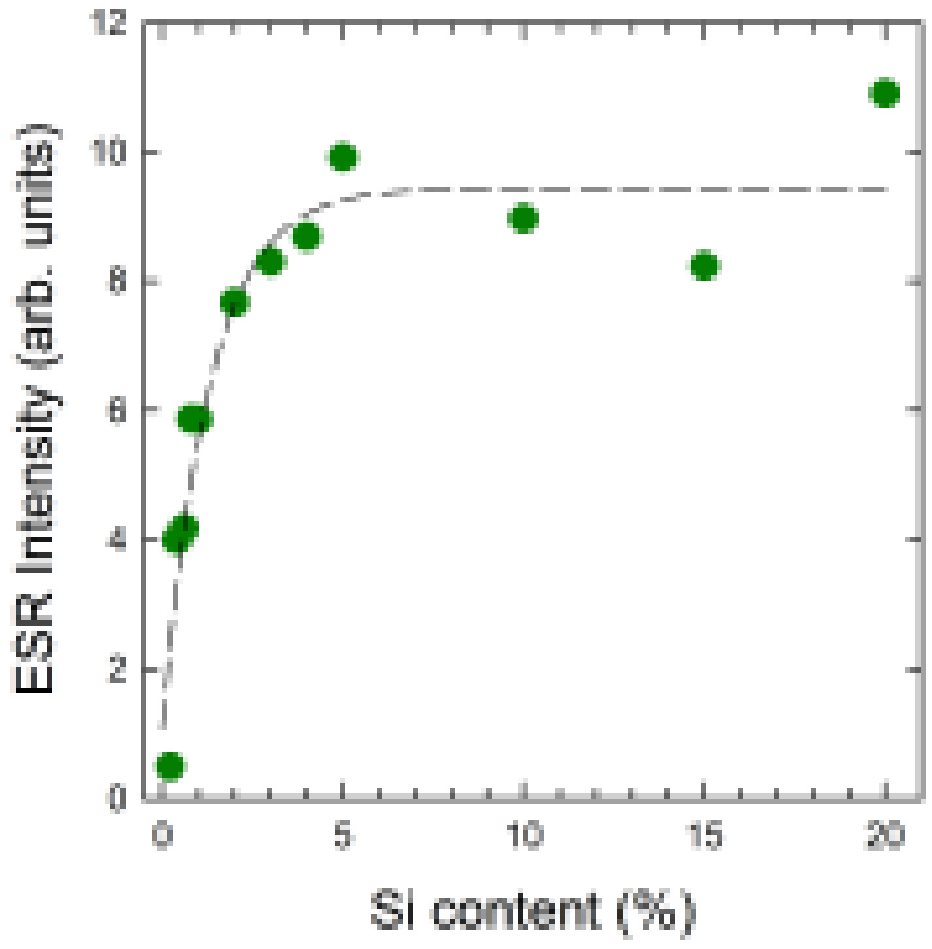

FIG.S6 The evolution of integrated ESR intensity for doped samples.